\documentclass[lettersize, journal]{IEEEtran}
\pdfoutput=1
\usepackage{amsmath, amsfonts}
\usepackage{algorithmic}
\usepackage{algorithm}
\usepackage{array}
\usepackage[caption=false, font=normalsize, labelfont=sf, textfont=sf]{subfig}
\usepackage[colorlinks, linkcolor=black, anchorcolor=black, citecolor=black]{hyperref}
\usepackage{textcomp}
\usepackage{stfloats}
\usepackage{color}
\usepackage{float}
\usepackage{url}
\usepackage{verbatim}
\usepackage{graphicx}
\usepackage[compress]{cite}
\usepackage{multirow}
\usepackage{bm}

\newcommand{\refsec}[1]{Section~\ref{#1}}
\newcommand{\reffig}[1]{Fig.~\ref{#1}}
\newcommand{\reftab}[1]{Table~\ref{#1}}
\newcommand{\refalg}[1]{Algorithm~\ref{#1}}

\newcommand{\etal}{\emph{et al}. }



\begin{document}
	
\title{Detection of False Data Injection Attacks in Smart\\ Grid: A Secure Federated Deep Learning Approach}

\author{Yang Li,~\IEEEmembership{Senior Member,~IEEE}, Xinhao Wei, Yuanzheng Li,~\IEEEmembership{Member,~IEEE},\\Zhaoyang Dong,~\IEEEmembership{Fellow,~IEEE}, and Mohammad Shahidehpour,~\IEEEmembership{Life Fellow,~IEEE}
	
\thanks{This work is supported by National Natural Science Foundation of China under Grant U2066208, the Natural Science Foundation of Jilin Province, China under Grant YDZJ202101ZYTS149, and Open Project of Key Laboratory of Modern Power System Simulation and Control and Renewable Energy Technology, Ministry of Education, Northeast Electric Power University under Grant MPSS2022-04.~\emph{(Corresponding author: Yuanzheng Li.)}}

\thanks{Y. Li, X. Wei are with the Key Laboratory of Modern Power System Simulation and Control and Renewable Energy Technology, Ministry of Education, Northeast Electric Power University, Jilin 132012, China (e-mail: liyang@neepu.edu.cn; 779058643@qq.com)}

\thanks{Y. Z. Li is with the School of Artifcial Intelligence and Automation, Huazhong University of Science and Technology, Wuhan 430074, China (e-mail: Yuanzheng\_Li@hust.edu.cn).}

\thanks{Z. Y. Dong is with the School of Electrical and Electronic Engineering, Nanyang Technological University, Singapore 639798 (e-mail: zy.dong@ntu.edu.sg).}

\thanks{Mohammad Shahidehpour is with the Electrical and Computer Engineering Department, Illinois Institute of Technology, Chicago, IL, 60616, USA. He is also a Research Professor in the ECE Department at the King Abdulaziz
University in Saudi Arabia (e-mail: ms@iit.edu).}}


\maketitle

\begin{abstract}
	As an important cyber-physical system (CPS), smart grid is highly vulnerable to cyber attacks. Amongst various types of attacks, false data injection attack (FDIA) proves to be one of the top-priority cyber-related issues and has received increasing attention in recent years. However, so far little attention has been paid to privacy preservation issues in the detection of FDIAs in smart grids. Inspired by federated learning, a FDIA detection method based on secure federated deep learning is proposed in this paper by combining Transformer, federated learning and Paillier cryptosystem. The Transformer, as a detector deployed in edge nodes, delves deep into the connection between individual electrical quantities by using its multi-head self-attention mechanism. By using federated learning framework, our approach utilizes the data from all nodes to collaboratively train a detection model while preserving data privacy by keeping the data locally during training. To improve the security of federated learning, a secure federated learning scheme is designed by combing Paillier cryptosystem with federated learning. Through extensive experiments on the IEEE 14-bus and 118-bus test systems, the effectiveness and superiority of the proposed method are verified. 
\end{abstract}


\section*{Abbreviation}
\leftline{
\begin{tabular}{ l l } 
	\centering
	 ~ & ~\\
	 FDIA & false data injection attack \\ 
	 CPS & cyber-physical system \\ 
	 ISO & independent system operator \\ 
	 BDD & bad data detection \\ 
	 DC & direct current \\ 
	 AC & alternating current \\
	 CNN & convolutional neural network \\
      GAN & generative adversarial network \\
\end{tabular}
}

\vspace{1cm}
\leftline{
\begin{tabular}{ l l }
	\centering
	 LSTM & long short-term memory \\
	 SecFed & secure federated learning \\
	 EMS & energy management system \\
	 GRU & gate recurrent unit \\
	 LSGAN & least square generative adversarial networks \\
	 SCADA & supervisory control and data acquisition \\
	 WLS & weighted least square \\
	 AMCNN & attention mechanism-based convolutional \\ 
	 & neural network \\
	 DG & distributed generation\\
	 MTD & moving target defense\\
	 & \\
\end{tabular}
}

\section{Introduction}
\IEEEPARstart{A}{s} power system is a critical infrastructure of a country, its safe operation is vital to national security. The automation of the entire power system is an important feature of the development of the electrical industry~\cite{1}. To achieve this goal, a modern power system is developing into a type of CPSs that is deeply integrated by sensing, communication, computing, decision-making, and control~\cite{2}. Due to the open communication environment and increasingly complex information-physical coupling of smart grid, cyber security has become an important factor affecting the secure operation of smart grids~\cite{3}. In December 2015, Ukrainian power grid suffered catastrophic consequences due to a massive blackout caused by a hacker attack~\cite{4}. As a new type of cyber attacks, FDIA seriously threatens the secure operation of smart grids, so it is crucial to establish an effective and efficient FDIA detection mechanism.

Up to date, many researchers have conducted research from the perspective of FDIA detection and achieved certain results~\cite{5, 6}. The study in~\cite{7} is one of the seminal studies investigating the vulnerability of DC state estimators about FDIA. And many studies on methods for DC state estimation have subsequently emerged, such as DC FDIA detetion by Kalman filtering~\cite{8}, network theory~\cite{9, 10}, and machine learning~\cite{11}. In~\cite{12}, a CNN-based method is proposed for detecting and locating FDIA in DC-model power system. In~\cite{13}, Deng~\etal put forward a countermeasure, named moving target defense, against FDIAs with limited meter information. In~\cite{14}, Deng~\etal analyze MTD against FDIAs on electrical grids. These methods show satisfactory performance in the detection of DC FDIA, but the DC model makes a large number of simplifications compared to the AC model, making it different from the AC model used in real life. Therefore some studies on AC state estimation of FDIA emerge~\cite{15, 16, 17}. In~\cite{18}, graph theory is introduced into FDIA detection work, based on that an FDIA detection method is proposed. J. J. Q. Yu~\etal\cite{19} use GRU and wavelet transform to detect FDIA in power system. Y. Zhang~\etal\cite{20} combine GAN with self-encoder for FDIA detection. X. Luo~\etal\cite{21} detect and locate FDIA in smart grid by interval observer. All these traditional methods are centralized detection methods. With the development of smart grids, power systems are becoming larger and larger and generating more and more data, the traditional centralized processing methods cannot cope with the explosive growth of data volume in smart grid. The above methods require all data from each node in the grid to be transmitted to a data center for inspection. Such centralized detection methods have the following problems: 1) these methods are reliant on a central work station which has powerful calculation and storage capabilities; 2) limited communication and storage resources make the grid less capable of processing data in real time, which can result in some critical operations of the grid not being executed in a timely manner; 3) the way that data is stored at the central workstation can easily lead to data breaches and cyber attacks; 4) most importantly, since one independent system operator (ISO) typically only has a limited number of attack samples and these samples are always related to highly-sensitive information, the owners are usually reluctant to share such attack instances considering privacy preservation. These issues make it a very challenging problem to train a data-driven FDIA detection model with desirable performance in a privacy-preserving manner for smart grids.

To solve the above problems, distributed detection methods emerges, and there are very few studies on distributed FDIA detection in the existing literature. Specifically, reference~\cite{22} proposes a FDIA detection method based on sub-grid-oriented microservice framework. Reference~\cite{23} proposes a distributed detection with adaptive sampling sequence, which can reduce communication overhead and improve detection efficiency while ensuring robustness; however, data fusion between neighboring sub-areas must be performed to train a wide-area cyber-attack detector in this approach. 

As a representative of distributed machine learning methods, federated learning has been successfully applied in industrial internet of things. For instance, reference~\cite{24} combines federated learning and AMCNN-LSTM to detect anomalies using data from time series. Reference~\cite{25} integrates federated learning and CNN-GRU and used it to detect cyber attacks in industrial internet of things. Reference~\cite{26} combines federated learning with LSGAN together and used to model renewable energy sources. Reference~\cite{27} integrates federated learning and attentive aggregation to do FDIA detection in industry 4.0. Unfortunately, federated learning has so far rarely been used for detecting stealthy FDIA in smart grids.

In order to bridge this gap, based on our existing work in~\cite{26}, a federated learning-based FDIA detection method is proposed in this paper by setting up a edge node detector at each node of power system to collect, store and detect data directly instead of an original central workstation, transforming traditional centralized detection methods into a distributed detection approach. By building a detector based on Transformer model, the proposed approach can fully extract data features via self-attention mechanism. The main contributions of this paper are as follows:
\begin{itemize}
	\item [1)] We propose a federated learning-based FDIA detection approach. By keeping the data locally during training, this approach protects the data privacy of each node in a power system while utilizing the data from all nodes to collaboratively train a detection model. Moreover, as the detection model is deployed on each node locally, the communication delays caused by transferring data between each node and dispatch center are avoided during online detection, thus enabling efficient data detection.
	\item [2)] We build a new Transformer-based FDIA detection model. By using its unique multi-head self-attention mechanism, the Transformer is able to delve deep into the connection between individual electrical quantities, thus effectively detecting FDIA with various intensities.
	\item [3)] We combine the Paillier cryptosystem with federated learning to build SecFed scheme. The cryptosystem encrypts the data exchanged by federated learning, preventing hackers from using the weights generated during training to deduce the original data information, which greatly improves the security of federated learning. To our knowledge, ours is the first report on leveraging secure federated learning for FDIA detection in smart grids.
	\item [4)] We perform extensive experiments to examine the effectiveness and superiority of the presented secure federated deep learning-based FDIA detection method. In addition, we also add noises on the measurement data to test the robustness of our method.
\end{itemize}

\section{Preliminary}
\subsection{Bad Data Detection}
Since measurement data in the power grid usually suffers from data incompleteness and data anomalies, state estimation is necessary to accurately and efficiently monitor state information to offer support for system security evaluation~\cite{28}. Existing bad data detection mechanisms for detecting state information in the EMS have the ability to resist some common less stealthy cyber attacks. However, there are still serious vulnerabilities in the bad data detection mechanism. Liu's team found that by following certain conditions, it is possible to construct spurious measurement data to be injected into SCADA without increasing the estimate residuals, which is stealthy to BDD mechanisms~\cite{7}.

As one of the basic analysis tools for EMS to achieve reliable monitoring of the power system, state estimation provides feedback to the system on the operating status of the network based on measurement data and network information. Currently, AC state estimation is commonly used in smart grids, which uses a nonlinear function between the measurement values and the system states. The corresponding model for state estimation is described as follows:

\begin{equation}
	\bm{z} = h(\bm{x})+\bm{e},
	\label{XX1}
\end{equation}
where $\bm{z}=\{z_1, z_2, \cdots, z_I\}$ is the measurement vector, which represents the power injection of each bus and the power flow of each branch; $\bm{x}=\{x_1, x_2, \cdots, x_J\}$ is the state vector, which represents the voltage amplitude and phase angle of each bus; $h(\bm{x})$ denotes the measurement function of $\bm{x}$; $\bm{e}=\{e_1, e_2, \cdots, e_I\}$ stands for the measurement noises; $I$ and $J$ denote the number of measurement variables and the number of state variables, respectively. 

Multiple approaches for solving the system state vectors $\bm{x}$ have been proposed in~\cite{30}. The most commonly used state estimate method in power systems is WLS. WLS solves for the estimate with the smallest objective function value as the optimal state result by using the sum of the squares of the differences between the measurement vector $\bm{z}$ and the estimated state vector $\bm{x}$ as the objective function, and the value of the estimated system state vector $\bm{x}'$ can be solved by 

\begin{equation}
	\bm{x}'=\underset {\bm{x}}{\text{arg min}}[\bm{z}-h(\bm{x})]^{T}\mathbf{Y}[\bm{z}-h(\bm{x})],
	\label{XX2}
\end{equation}
\noindent where $\mathbf{Y}$ is a diagonal matrix, and each of its elements is equal to the inverse of the respective measurement accuracy.

During the whole process of remote terminal measurement data from being collected in the field to being transmitted to the database of the power control center, each step may be subject to random disturbances and generate errors, such as sensor offset, interference in the communication process, and human error. These errors can cause some measurement data to deviate from the real value and make it significantly different from the normal data. This kind of measurement data is called bad data. Bad data with large errors, when subjected to the state estimation process, can cause the calculated deviations between the calculated state estimate and the real condition, which seriously affects the control center operator's judgment of the system state. Researchers have adopted a series of methods to detect, process, and eliminate bad data. Most of the commonly used methods are based on the residual test principle. The residual $\bm{r}$ is defined as
\begin{equation}
	{\left\| \bm{r} \right\|_2}=\Vert{\bm{z}-h(\bm{x}')\Vert_2}.
	\label{XX3}
\end{equation}

Assuming that all state vectors are independent mutually and the random measurement error vectors obey the Gaussian distribution with zero mean, therefore it is attested that ${\Vert{\bm{r}}\Vert_2}^2$ obeys the chi-square distribution. In order to detect whether there is bad data in the measurements, we compare euclidean norm of residual with the threshold value $\tau$~\cite{29}. The specific inequality is shown as below:
\begin{equation}
	\Vert{\bm{r}}\Vert_2>\tau.
\end{equation}

\subsection{False Data Injection Attack}
Changing some state vectors by manipulating a set of measurement vectors is the basic principle of constructing a FDIA on an AC power system~\cite{18}. For instance, if an attacker intends to change the real power on bus $i$, he ought to make a series of attack vectors so that the estimated state vectors do not differ from the actual state values. Due to the subjection to the power flow equations, all the measurement vectors of the system have to follow the change of the state vectors. The power flow equations are as follows:

\begin{equation}
	{P_i} = \sum\limits_{j = 1}^{{N_i}} {{V_i}{V_j}({G_{ij}}\cos{\theta_{ij}} + } {B_{ij}}\sin{\theta_{ij}}),
	\label{XX4}
\end{equation}
\begin{equation}
	{Q_i} = \sum\limits_{j = 1}^{{N_i}} {{V_i}{V_j}({G_{ij}}\sin{\theta_{ij}} - } {B_{ij}}\cos{\theta_{ij}}),
	\label{XX5}
\end{equation}
\begin{equation}
	{P_{ij}} = V_i^2({g_{si}} + {g_{ij}}) - {V_i}{V_j}({g_{ij}}\cos{\theta_{ij}} + {b_{ij}}\sin{\theta _{ij}}),
	\label{XX6}
\end{equation}
\begin{equation}
	{Q_{ij}} =  - V_i^2({b_{si}} + {b_{ij}}) - {V_i}{V_j}({g_{ij}}\sin{\theta_{ij}} - {b_{ij}}\cos{\theta _{ij}}),
	\label{XX7}
\end{equation}
where ${P_i}$ represents the real power injections at bus \emph{i}, ${Q_i}$ denotes reactive power injections at bus \emph{i}. ${P_{ij}}$ stands for the real power flow from bus \emph{i} to bus \emph{j}, ${Q_{ij}}$ represents the reactive power flow from bus \emph{i} to bus \emph{j}. $V_i(V_j)$ is the voltage magnitude at bus \emph{i} (\emph{j}), $\theta_{i}(\theta_{j})$ represents the phase angle at bus \emph{i} (\emph{j}). $G_{ij}+jB_{ij}$ is the \emph{ij}th element of the bus admittance matrix, ${g_{ij}}+j{b_{ij}}$ is the admittance of the branch between buses \emph{i} and \emph{j}, ${g_{si}}+j{b_{si}}$ is the admittance of the shunt branch connected at bus \emph{i}, $N_i$ indicates the number of buses connected to bus \emph{i}, and ${\theta_{ij}}={\theta_i}-{\theta_j}$.

From (5)-(8), we can see that any change in the state value affects the change in the measurement values correlating to it. For creating a stealthy FDIA in power system, a set of measurements needs to be changed.

A. Abur~\etal\cite{30} state that in order to make attack vectors stealthy, an attack vector $\bm{\alpha}$ could be constructed by
\begin{equation}
	\bm{\alpha} = h(\bm{x}' + \bm{l})-h(\bm{x}'),
	\label{XX8}
\end{equation}
where $\bm{l}$ denotes the vector of deviations of state quantities after being attacked. By adding the attack vector $\bm{\alpha}$ to the real measurement vector, the measurement vector of the attack will change to $\bm{z}_\alpha= \bm{z}+\bm{\alpha}$, and accordingly the state vector of the attacked system will become $\bm{x}'_\alpha=\bm{x}+\bm{l}$. When the attack vector is constructed using the above method, the residuals after being attacked will become as shown below:

\begin{equation}
	\begin{aligned}
	{\left\| {{\bm{r}_\alpha}} \right\|_2} &= {\left\| {{\bm{z}_\alpha} - h({{\bm{x}'}_\alpha})} \right\|_2} = {\left\| {\bm{z} + \bm{\alpha} - h(\bm{x}' + \bm{l})} \right\|_2}
	\\&= {\left\| {\bm{z} + h(\bm{x}' + \bm{l}) - h(\bm{x}') - h(\bm{x}' + \bm{l})} \right\|_2} 
	\\&= {\left\| \bm{r} \right\|_2}.
	\label{10}
	\end{aligned}
\end{equation}

From (10), it can be clearly seen that the residual values do not change and therefore the bad data detector can not detect the attack. In this way, a stealthy FDIA is built on the vulnerability of conventional bad data detection. Note that, this attack is able to modify the original measurements while bypassing the bad data detection in power systems, as it keeps the 2-norm of the compromised data unchanged in such cases.

\section{Proposed Method}
Firstly, the workflow of the proposed SecFed scheme is described. Then, the developed transformer-based FDIA detection model is presented. Finally, the federated learning-based framework is demonstrated.

\subsection{Proposed SecFed Scheme}
\begin{figure}[H]
	\centering
	\includegraphics[width=3.5in]{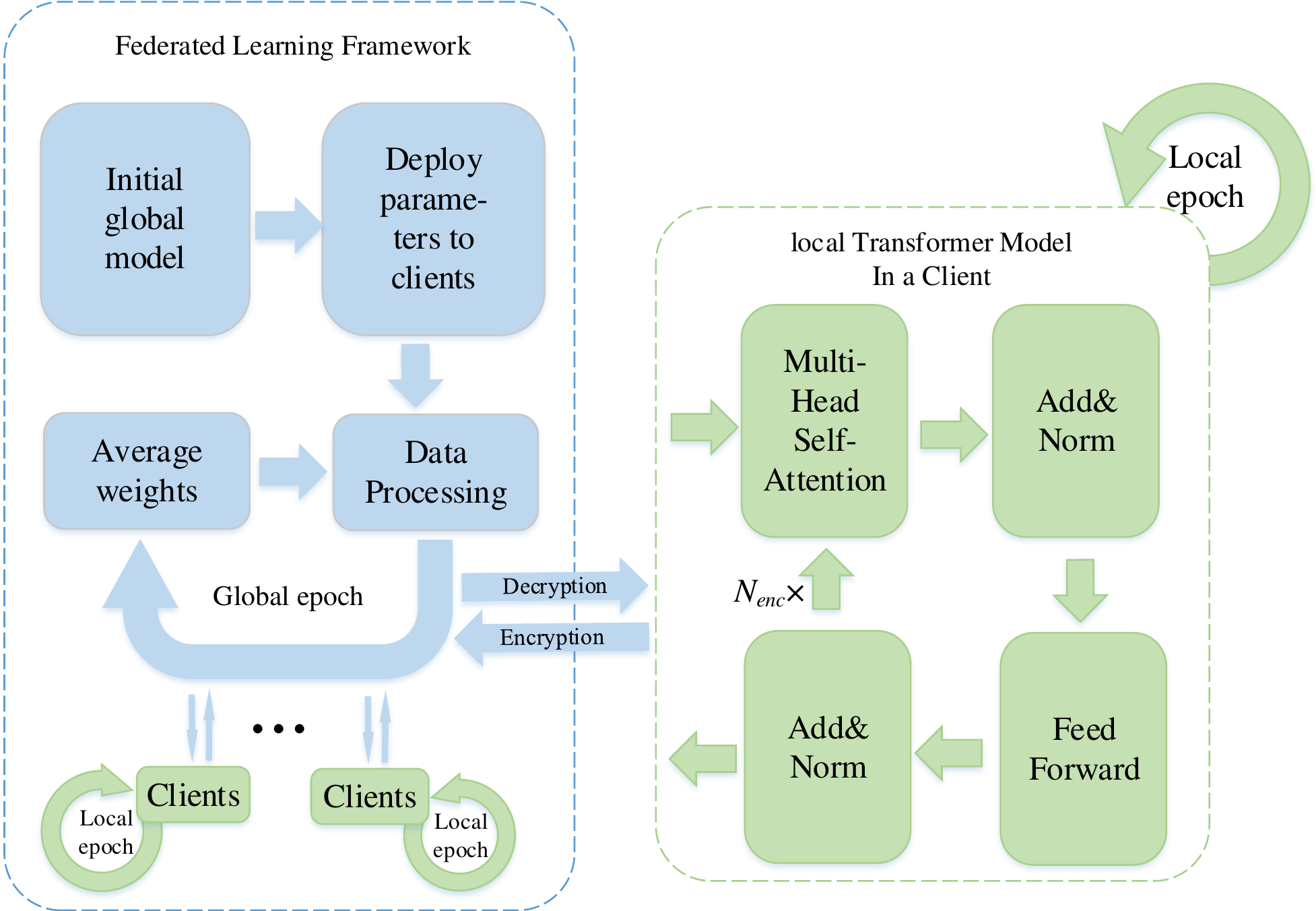}
	\caption{Flow Chart of the SecFed-Transformer.}
	\label{fig1}
\end{figure}

The key idea of the SecFed scheme is that multi-clients co-train a FDIA detection model while keeping the data locally. This scheme is a combination of the federated learning and the Paillier cryptosystem. \reffig{fig1} and~\refalg{alg1} show the flowchart of the SecFed-Transformer and the details of the scheme, respectively. To be specific, the SecFed scheme consists of the following five steps:

1) Initialization: After all clients are enumerated by the trustee, the model parameters are initialized for each client, such as the number of encoder blocks $N_{enc}$, the number of communication rounds $R$, the number of local epochs $E$, size of minibatch $B$, learn rate $lr$, loss function $L$, initial weight ${w_0}$, the communication round index $r$, the time step $e$, the first moment vector $f$, the second moment vector $s$, exponential decay rates for moment estimates ${\alpha _1}$ and ${\alpha _2}$, a small constant used for numerical stabilization $\varepsilon$. Generate public key ${PK}=(n, g)$ and private key ${CK}(\lambda, \mu)$ based on the Paillier cryptosystem through ${KeyGeneration}$ (see Appendix for the details).

2) Training of local models deployed on each client: After initialization, each client locally trains a Transformer-based FDIA detection model using their respective local data.

3) Encryption of model weights: When the training of each client's local Transformer model is completed, each client extracts and encrypts the respective model weights by ${Encryption}(w, {PK})$, and then uploads the encrypted model weights to the cloud.

4) Aggregation of model weights: According to the received encrypted model weights from each client, the cloud server aggregates them through ${Aggregation}({c_{td}})$, and then the aggregated ciphertext is sent back to each client.

5) Update of local models: By using the method of ${Decryption}({a_{d}}, {CK})$, the ciphertext $c$ is decrypted, and each client obtains the updated model weights. And then, the local deep learning model is updated.

To facilitate the analysis, we make the following three assumptions: i) all clients use the same initial global framework to train their Transformer models; ii) all local models have the same hyperparameters and the same optimization algorithms; iii) the computing power of each client is similar. It should be noted that this paper does not consider dropped packets in the communication process, while  realistic situations need to consider these issues. Referring to our previous study~\cite{26}, we set the federated learning related parameters. It is noted that the cloud server itself does not train the model, and the knowledge of the weights of the global model is derived by aggregating the weights of each local model.

\begin{algorithm}
	\caption{Secure Federated Learning(SecFed).}
		\begin{algorithmic}[1]
		\STATE \textbf{Input}: The data resources of each client $\{{D_t} \mid t\in T\}$, the number of communication rounds \emph{R}, the number of local epochs \emph{E}, the size of minibatch \emph{B}, learning rate \emph{lr}, loss function \emph{L}.
		\STATE \textbf{Output}: The SecFed-Transformer model.
		\STATE \textbf{Initialization}:
		\STATE \hspace{0.2cm}a). The trustee generates private key and public key
		\\ \hspace{0.2cm}$\{{CK}, {PK}\}$ by ${KeyGeneration}$ and sends them to each 
		\\ \hspace{0.2cm}client.
		\STATE \hspace{0.2cm}b). Initialize the ${w_0}$.
		\STATE \hspace{0.2cm}c). Initialize the \emph{r} = 0.
		\STATE \textbf{Procedure}:
		\STATE \textbf{for} $r \leq R$ \textbf{do}
		\STATE \hspace{0.2cm}\textbf{(I). For each client:}
		\STATE \hspace{0.2cm}\textbf{Initialize} \emph{e} = 1
		\STATE \hspace{0.2cm}\textbf{for} each round $e \leq E$ \textbf{do}
		\STATE \hspace{0.4cm}$B \gets$ (split ${D_t}$ into minibatch of size $B$)
		\STATE \hspace{0.4cm}\textbf{for} each minibatch of data resource \textbf{do}
		\STATE \hspace{0.6cm}Initialize $f = 0, s = 0$, ${\alpha _1}$ = 0.9, ${\alpha _2}$ = 0.999, \\ \hspace{0.6cm}$\varepsilon=10^{-8}$, respectively;
		\STATE \hspace{0.6cm}a). Compute the gradient by $g \gets \nabla {w_{t, r}L}$.
		\STATE \hspace{0.6cm}b). Renew the biased first moment estimation by \\ \hspace{0.6cm}$f \gets {\alpha _1}f + (1 - {\alpha _1})g$.
		\STATE \hspace{0.6cm}c). Renew the biased second raw moment estimation \\ \hspace{0.6cm} by $s \gets {\alpha _2}s + (1 - {\alpha _2}){g^2}$.
		\STATE \hspace{0.6cm}d). Calculate the bias-corrected first moment estimat-\\  \hspace{0.6cm} ion by $\hat f \gets \frac{f}{{1 - \alpha _1^e}}$.
		\STATE \hspace{0.6cm}e). Calculate the bias-corrected second raw moment \\ \hspace{0.6cm} estimation by $\hat s \gets \frac{s}{1 - \alpha _2^e}$.
		\STATE \hspace{0.6cm}f). Renew the model weights by \\ \hspace{0.6cm} ${w_{t, r}} \gets {w_{t, r}}-lr\frac{{\hat f}}{{\sqrt {\hat s}+ \varepsilon}}$.
		\STATE \hspace{0.4cm}\textbf{end}
		\STATE \hspace{0.4cm}\textbf{return}${w_{t, r}}$
		\STATE \hspace{0.4cm}\textbf{for} each ciphertext ${c_{td}}$ \textbf{do}
		\STATE \hspace{0.6cm}Compute the ciphertext by
		\STATE \hspace{0.6cm}${c_{td}} \gets {Encryption}({w_{t, r}}, {PK}$);
		\STATE \hspace{0.4cm}\textbf{end}
		\STATE \hspace{0.4cm}\textbf{(II). For cloud server:}
		\STATE \hspace{0.4cm}\textbf{for} each ciphertext ${c_{td}}$ \textbf{do}
		\STATE \hspace{0.6cm}Aggregate the ciphertext by 
		\STATE \hspace{0.6cm}${a_{d}} \gets {Aggregation}({c_{td}}$);
		\STATE \hspace{0.4cm}\textbf{end}
		\STATE \hspace{0.4cm}The cloud server distributes the aggregated ciphertext \\ \hspace{0.4cm} ${a_d}$  to all clients.
		\STATE \hspace{0.4cm}\textbf{(III). For each client:}
		\STATE \hspace{0.4cm}\textbf{If} $r < R$
		\STATE \hspace{0.6cm}\textbf{for} each aggregated ciphertext ${a_d}$ \textbf{do}
		\STATE \hspace{0.8cm}Decrypt the ciphertext by
		\STATE \hspace{0.8cm}${w_{t, r}} \gets {Decryption}({a_d}, {CK}$);
		\STATE \hspace{0.6cm}\textbf{end}
		\STATE \hspace{0.6cm}Each client upload its local model using the renewed \\ \hspace{0.6cm} weights ${w_{i, r+1}}$
		\STATE \hspace{0.6cm} $r \gets r+1$
		\STATE \hspace{0.4cm}\textbf{Else}
		\STATE \hspace{0.6cm}\textbf{Break}
		\STATE \textbf{end}
		\STATE \textbf{return} The comprehensive deep learning model.
		\end{algorithmic}
		\label{alg1}
\end{algorithm}

\subsection{Distributed Detector Based on Transformer}
\begin{figure}[h]
	\centering
	\includegraphics[width=3.5in]{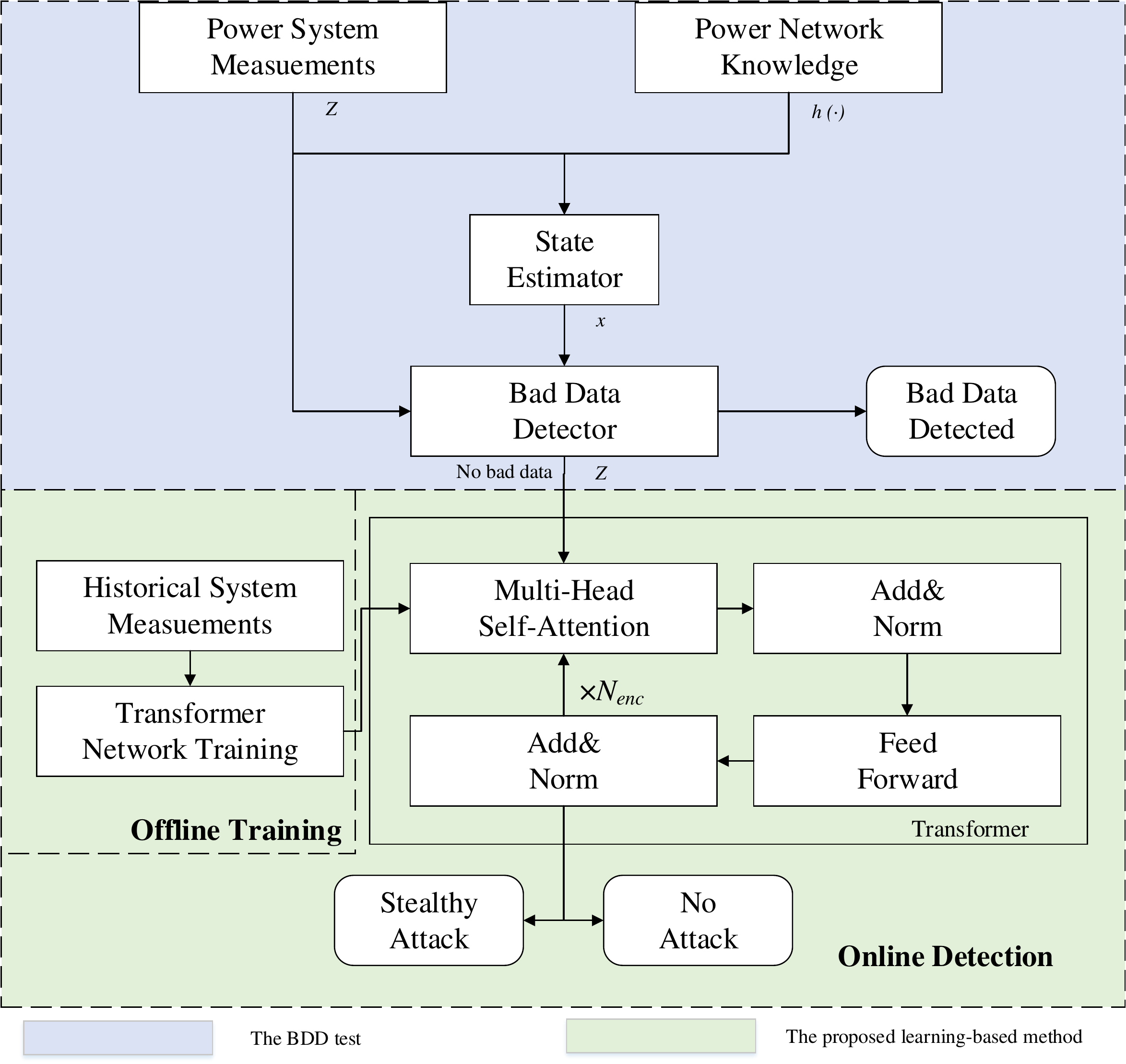}
	\caption{Proposed Transformer-based FDIA detector.}
	\label{fig_2}
\end{figure}

As a cutting-edge deep learning model, Transformer firstly proposed in 2017 has shown great success in the area of language recognition in recent years~\cite{31}. Unlike most of the current major sequence transduction models that are based on complex neural networks recursion or convolution of neural networks, Transformer eschews recursion and convolution and forms its network structure through attention mechanisms and neural networks. In addition, traditional deep learning models perform feature extraction based on the sequential ordering of data~\cite{zhangmeng}, while Transformer focuses more on the global data feature relationships, which makes it more suitable for dealing with FDIA detection issues. In this section, a deep learning model based on Transformer is presented to detect stealthy FDIA in smart grid, and a diagram of the proposed detector is depicted in~\reffig{fig_2}. 

1) The input and output of the detector: In this work, active and reactive power injections of nodes and power flows of branches are adopted as the input of the detection model, while the judgment of whether a FDIA occurs as the output. We formulate the problem for detecting stealthy FDIA represented by the attack vector $\bm{\alpha}$ as a binary classification problem with the detection indicator $\beta$.
\begin{equation}
	\beta= \begin{cases}0, & \text{Normal data} \\ 1, & \text{Compromised data.}\end{cases}
\end{equation}

In the Transformer-based detection model, the measurement vector $\bm{z}$, comprising power injections and power flows, is gathered as the inputs of Transformer, which is labeled with $\beta=1$ or $\beta=0$ . In comparison to the traditional WLS detection approach, our detection method is totally data-driven and does not need the model and parameters of the analyzed system, i.e., $h(\bm{x})$. In addition, we train the Transformer model to deeply extract attack features from normal and compromised data, and thereby detect the presence of stealthy FDIAs.

2) Transformer architecture: A Transformer model mainly consists of a positional encoding, encoder blocks and a sigmoid layer, where a encoder block includes a multi-head self-attention module, Add\&Norm layers and feedforward neural networks.

a) Positional encoding: In order to learn the intrinsic laws obeyed by the feature quantities of different positions under normal and compromised conditions, the Transformer must record the position information of each feature quantity in sample sets. The data is first processed for recording the position information of each feature quantity in a data sample by positional encoding, which can be described as 

\begin{equation}
	\begin{aligned}
		\mathrm{PE}_{{(\phi, 2k)}} = \rm{sin}(\phi/{10000^{2k/d}}),
		\label{XX10}
	\end{aligned}
\end{equation}
\begin{equation}
	\begin{aligned}
		\mathrm{PE}_{{(\phi, 2k+1)}} = \rm{cos}(\phi/{10000^{2k/d}}),
		\label{XX11}
	\end{aligned}
\end{equation}
where $\mathrm{PE}(\cdot)$ is the function of positional encoding operations, $\phi$ denotes the position of each sample in sample sets, $d$ represents the dimension, 2\emph{k} denotes the even dimension, and 2\emph{k}+1 denotes the odd dimension (\emph{k} is a natural number, \emph{$2k\leq d, 2k+1\leq d$}).


b) Multi-head self-attention module: This module uses the self-attention mechanism to dig out the potential relationships in the electrical quantities of the power system to distinguish the true values from the false ones. The multi-head self-attention module consists of several self-attention modules.

 Self-attention uses matrices \textbf{Q}(query), \textbf{K}(key), and \textbf{V}(value) in its calculation. In this study, the self-attention receives the output of positional encoding or the output of the previous encoder block as its input. \textbf{Q, K, V} are obtained by multiplying the input of self-attention by the linear matrix \textbf{WQ}, \textbf{WK}, \textbf{WV}, respectively. After obtaining the matrices \textbf{Q, K, V}, the output of self-attention can be calculated as the following formula:
\begin{equation}
	\begin{aligned}
		\mathrm{Attention}(\mathbf{Q},\mathbf{K},\mathbf{V}) = \mathrm{softmax}(\frac{{\mathbf{Q}{\mathbf{K}^T}}}{{\sqrt{{d_k}}}})\mathbf{V},
		\label{XX12}
	\end{aligned}
\end{equation}
\noindent where ${d_k}$ is the number of columns of matrices \textbf{Q} and \textbf{K}.
The obtained self-attention matrix reflects the degree of correlation between each electrical quantity. 

\indent c) Add\&Norm layer: This layer is set after the multi-head self-attention module or feedforward neural network module. The calculation of Add\&Norm layer consists of two operations: Add and Norm. Here, Add stands for residual connection to prevent network degradation, and Norm represents layer normalization to normalize the activation values of each layer.  The specific operations in this layer are given as follows:

\begin{equation}
	\begin{aligned}
		\mathrm{LayerNorm}(\mathbf{X} + \mathrm{MultiHeadAttention}(\mathbf{X})),
		\label{XX13}
	\end{aligned}
\end{equation}

\begin{equation}
\begin{aligned}
	\mathrm{LayerNorm}(\mathbf{X} + \mathrm{FeedForward}(\mathbf{X}))\label{XX14},
\end{aligned}
\end{equation}


\noindent where \textbf{X} is the input of multi-head self-attention module or feedforward neural network module. The Add operation refers to $\mathbf{X} + \mathrm{MultiHeadAttention}(\mathbf{X})$ or $\mathbf{X} + \mathrm{FeedForward}(\mathbf{X})$, which is used to solve the problem of multi-layer network training such as gradient dispersion and gradient explosion; while the Norm operation refers to layer normalization, which can speed up the convergence process by converting the input of each layer to have the same mean-variance.

\indent d) Feedforward neural network module: This module consists of two parts, a fully connected layer in the front as well as a dropout layer in the back. The fully connected layer corresponds to the following equation:

\begin{equation}
	\begin{aligned}
		\max(0, \mathbf{X}{W_1} + {b_1}){W_2} + {b_2},
		\label{XX15}
	\end{aligned}
\end{equation}
where \textbf{X} is the input, and $\mathrm{FeedForward}$ ends up with an output matrix whose dimensions are consistent with \textbf{X}.

And then, we apply dropout to the output of feedforward neural network module. The role of the dropout layer is to prevent overfitting of the model.

e) Sigmoid layer: This layer consists of two fully connected layers in series, and sigmoid function is used as the activation function of the last fully connected layer. The Sigmoid layer is used to map the output of the Transformer to a classification result.

\subsection{Proposed Framework Based on Federated Learning}
Federated learning, initially studied and proposed by the Google team~\cite{32}, is a machine learning framework that accomplishes multi-client collaboratively modeling without data sharing. This paper assumes that there are $T$ clients (e.g., edge devices) and each client has a separate piece of data. A traditional centralized training approach builds a large dataset by collecting data from all clients in a central workstation, which trains the large dataset to obtain a global model. With the centralized approach, each client exposes its data to each workstation and other clients, which makes it very vulnerable to data leakage. In contrast, federated learning simply uploads the weights of each client to a central server for aggregation, with the original data for each client stored locally and not exchanged or transferred, and the weights are aggregated and updated to achieve learning goals. Although federated learning does not require original data in data sharing, participants still need to upload model weights to construct a federated model, which are essentially a mapping of raw data, and cloud servers or hackers can invert the data of each client through model weights. To address this problem, we introduce the Paillier cryptosystem~\cite{33}, a homomorphic encryption method, to protect the weights of the models in federated learning. The specific steps of the Paillier cryptosystem are shown in Appendix.

Combining the Paillier cryptosystem with traditional federated learning, we propose a secure federated learning approach for communication over multiple edge devices for collaborative training of multiple models. As shown in~\reffig{fig_3}, it is generally composed of the following components:
\begin{figure}[!t]
	\centering
	\includegraphics[width=3.5 in]{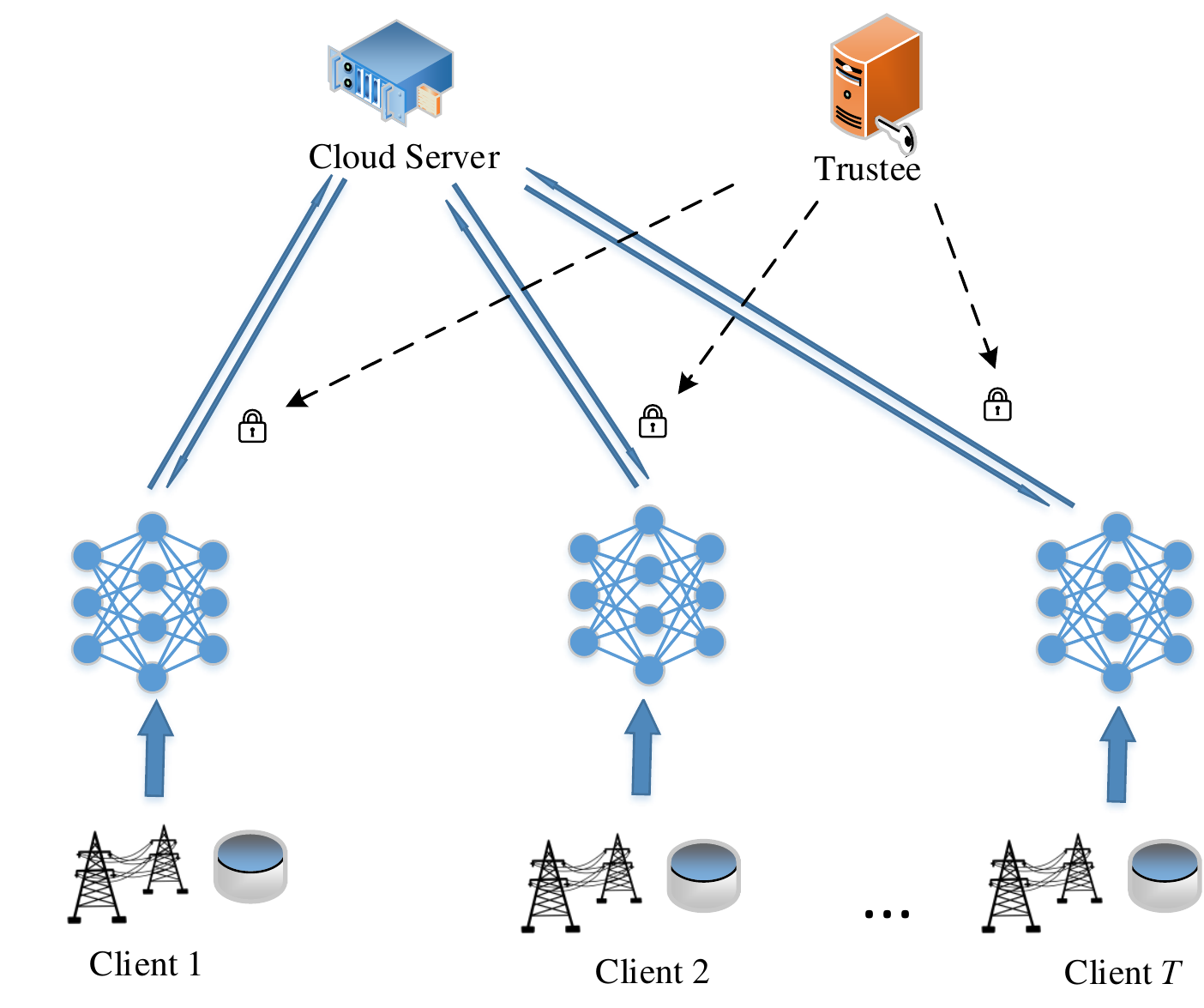}
	\caption{The structure of server-client-trustee in SecFed}
	\label{fig_3}
\end{figure}

\begin{itemize}
	\item Cloud server: A cloud server is usually a cloud aggregator which has powerful computing power and abundant computing resources. The main functions of cloud aggregator are as follows: (1) initializes the global model before the first run of the global model; (2) sends the global model to all clients after each communication round; (3) aggregates the weights which are uploaded by the clients until the model converges.
	\item Clients: Each client is responsible for modeling a Transformer-based local FDIA detector, which is based on its own gathered measurement data (time-section electrical data collected by SCADA at each node) on behalf of each client, and helps to update the weights of the detector by repeatedly interoperating with the cloud server until the detector converges. Note that, the presented method is also suitable for WAMS data or mixed measurement data of WAMS and SCADA, besides SCADA data.
	\item Trustee: Based on the Pillier cryptosystem, trustee generates public keys as well as private keys, and uses them to encrypt the weights which are uploaded by each client to the server, and decrypt the weights distributed by the server to each client.
\end{itemize}

\section{Example Analysis}
In this section, we conduct a large number of experiments to evaluate the performance of SecFed-Transformer method. In~\refsec{Data Generation}, we describe the data generation. In~\refsec{Experiment Settings}, we introduce the implementation details and hyperparameter settings. In~\refsec{Detection Results}, we conduct numerous of experiments for comparing SecFed-Transformer with some common deep learning detection models including the CNN and LSTM, under our SecFed scheme, and use different metrics to compare the proposed approach with traditional algorithms in detail. Finally, in~\refsec{The Effect of Measurement Noises}, the robustness of the proposed model is illustrated by observing the accuracy under different noises.

\subsection{Data Generation}\label{Data Generation}
1) Normal data: Following the approach in~\cite{12}, we firstly generate uncompromised data which is based on the IEEE 14-bus system and IEEE 118-bus system. Specifically, we generate a set of data that has a mean equal to the base load and a variance of one-tenth of the base load.

2) Compromised data: Since unstealthy FDIAs can be detected by the BDD, we only construct stealthy FDIAs. We generate compromised data under strong and weak attacks which is performed on the IEEE 14-bus system and IEEE 118-bus system.

According to~\cite{19}, it is stated that all generated FDIA samples are classified into 3 categories based on the “strength” of attacks: a) Weak attacks: The ratio of the mean of the power injection deviations in $c$ to $x$ is smaller than 10\%, or the ratio of the mean of the voltage magnitude deviations to nominal value is smaller than 5\%, or the mean of the voltage angle deviations is smaller than $2^{\circ}$; b) Strong attacks: The means of the deviations of the above-mentioned variables are respectively greater than 30\%, 10\%, and $5^{\circ}$; c) Medium attacks: FDIA samples not belonging to the above two attacks.

3) Training and test datasets settings: In each attack case, we generate a training set containing 10,000 samples of normal data and 10,000 samples of compromised data, and a test set containing 1,000 samples of normal data and 1,000 samples of compromised data. In addition, it should be noted that we choose horizontal federated learning, which requires the data features of each client to be consistent, so we choose the bus data of each node and a branch data connected to the bus for each client data. Therefore, each data set consists of power injections of all buses, as well as power flow of the selected branch.

\begin{table*}[hbp]
	\centering
	\caption{Data Comparison of Two Different Nodes of The IEEE 14-Bus system \\with Different Number of Communication Rounds (Under Weak Attacks)}
	\label{table1}
	\renewcommand{\arraystretch}{1.4}
	\begin{tabular}{ccccccccccc}
		\hline
		\multirow{2}{*}{Bus}         & \multirow{2}{*}{$R$}         & \multicolumn{3}{c}{CNN}                       & \multicolumn{3}{c}{LSTM}                      & \multicolumn{3}{c}{The Proposed Method}       \\\cline{3-11}
		                           &                            & Precision           & Recall           & F1-score            & Precision           & Recall           & F1-score            & Precision           & Recall           & F1-score            \\
		\hline
		\multirow{4}{*}{2}         & 0                          & 0.9841             & 0.9783        & 0.9812         & 0.9964        & 0.9888        & 0.9926        & 0.9991        & 0.9940         & 0.9965        
		\\
		                           & 2                          & 0.9991             & 0.9816        & 0.9903        & 0.9962        & 0.9894        & 0.9928        & 0.9985        & 0.9944        & 0.9964        
		                           \\
		                           & 4                          & 0.9991             & 0.9834        & 0.9912        & 0.9977        & 0.9887        & 0.9932        & 0.9995        & 0.9936        & 0.9966        
		                           \\
		                           & 6                          & 0.9991             & 0.9841        & 0.9915         & 0.9976        & 0.9886        & 0.9931        & 0.9995        & 0.9938        & 0.9966        \\
		\hline
		\multirow{4}{*}{3}         & 0                          & 0.5537        & 0.4380         & 0.4891        & 0.5919        & 0.4947        & 0.5389        & 0.9932        & 0.8188        & 0.8976        
		\\
		                           & 2                          & 0.5467        & 0.5091        & 0.5272        & 0.5893        & 0.4611        & 0.5174        & 0.9830          & 0.8725        & 0.9245        
		                           \\
		                           & 4                          & 0.5539        & 0.5019        & 0.5019        & 0.5826        & 0.5149        & 0.5467        & 0.9850         & 0.8779        & 0.9255        
		                           \\
		                           & 6                          & 0.5577        & 0.5180         & 0.5371        & 0.6336        & 0.6088        & 0.6210         & 0.9855        & 0.8883        & 0.9344  \\
		\hline                        
	\end{tabular}
\end{table*}

\begin{center}
	\begin{table*}[ht]
	\centering
	\caption{Data Comparison of Two Different Nodes of the IEEE 14-Bus System \\with Different Number of Communication Rounds (Under Strong Attacks)}
	\label{table2}
	\renewcommand{\arraystretch}{1.4}
	\begin{tabular}{ccccccccccc}
		\hline
		\multirow{2}{*}{Bus}         & \multirow{2}{*}{$R$}         & \multicolumn{3}{c}{CNN}                         & \multicolumn{3}{c}{LSTM}                      & \multicolumn{3}{c}{The Proposed Method}                      \\\cline{3-11}
		                           &                            & Precision           & Recall           & F1-score            & Precision           & Recall           & F1-score            & Precision           & Recall           & F1-score            \\
		\hline
		\multirow{4}{*}{2}         & 0                          & 1.0000              & 1.0000              & 1.0000             & 1.0000             & 0.9766        & 0.9981        & 1.0000             & 1.0000             & 1.0000             \\
		                           & 2                          & 1.0000              & 1.0000              & 1.0000             & 1.0000             & 0.9952        & 0.9976        & 1.0000             & 1.0000             & 1.0000             \\
		                           & 4                          & 1.0000              & 1.0000              & 1.0000             & 0.9997        & 1.000             & 0.9999        & 1.0000             & 1.0000             & 0.9990         \\
		                           & 6                          & 1.0000              & 1.0000              & 1.0000             & 1.0000             & 1.0000             & 1.0000             & 1.0000             & 1.0000             & 1.0000             \\
		\hline
		\multirow{4}{*}{3}         & 0                          & 0.7547         & 0.6859         & 0.7210         & 0.8591        & 0.6913        & 0.7661        & 1.0000             & 0.9257        & 0.9614        \\
		                           & 2                          & 0.7741         & 0.7044         & 0.7376        & 0.9329        & 0.8199        & 0.8727        & 1.0000             & 0.9998        & 0.9999        \\
		                           & 4                          & 0.7779         & 0.7096         & 0.7422        & 0.9837        & 0.8642        & 0.9200          & 0.9999        & 0.9999        & 0.9990         \\
		                           & 6                          & 0.7703         & 0.7054         & 0.7364        & 0.9768        & 0.9019        & 0.9379        & 1.0000             & 1.0000             & 1.0000            \\
		\hline
	\end{tabular}
	\end{table*}
\end{center}
\vspace{-1cm}
\subsection{Experiment Settings}\label{Experiment Settings}
1) Environmental setup: All simulations are performed on a machine with Intel i9-10900k CPU, GTX3090 GPU, and 32 GB RAM. Data generation is performed in Matlab using MATPOWER~\cite{matpower}, and the federated deep learning component is run in Python based on Tensorflow 2.5.

2) Hyperparameter settings: Regarding Transformer, the learning rate is selected as 0.0001, the number of encoder blocks is 3, the number of local epochs under strong attacks is set to 400, and the number of local epochs under weak attacks is 1000, the minibatch is set to 128, the Adam optimizer is selected as the optimizer, and the Binary Cross-Entropy  is chosen as the loss function. As for federated learning, the maximum number of communication rounds is 9. 

\begin{center}
	\begin{table*}[ht]
	\centering
	\caption{Data Comparison of Two Different Nodes of the IEEE 118-Bus System \\with Different Number of Communication Rounds (Under Weak Attacks)}
	\label{table3}
	\renewcommand{\arraystretch}{1.4}
	\begin{tabular}{ccccccccccc}
		\hline
		\multirow{2}{*}{Bus}         & \multirow{2}{*}{$R$}         & \multicolumn{3}{c}{CNN}                         & \multicolumn{3}{c}{LSTM}                      & \multicolumn{3}{c}{The Proposed Method}                      \\\cline{3-11}
		                           &                            & Precision           & Recall           & F1-score            & Precision           & Recall           & F1-score            & Precision           & Recall           & F1-score            \\
		\hline
		\multirow{4}{*}{55}         & 0                          & 0.6071         & 0.6578         & 0.6314         & 0.6157         & 0.6392        & 0.6273        & 0.8257        & 0.8206        & 0.8231        \\
		                            & 2                          & 0.8586         & 0.7387         & 0.7941         & 0.7341         & 0.6926        & 0.7127        & 0.9944        & 0.9279        & 0.9600          \\
		                            & 4                          & 0.8617         & 0.7415         & 0.7971         & 0.9201         & 0.7716        & 0.8393        & 0.9997        & 0.9355        & 0.9666        \\
		                            & 6                          & 0.8556         & 0.7342         & 0.7903         & 0.9492         & 0.7698        & 0.8501        & 0.9997        & 0.9355        & 0.9666        \\
		\hline
		\multirow{4}{*}{87}         & 0                          & 0.5845         & 0.5526         & 0.5681         & 0.7828         & 0.6718        & 0.7231        & 0.8276        & 0.7032        & 0.7603        \\
		                            & 2                          & 0.5826         & 0.5695         & 0.5758         & 0.9394         & 0.7903        & 0.8584        & 0.9844             & 0.8358        & 0.9037        \\
		                            & 4                          & 0.5824         & 0.5669         & 0.5745         & 0.9402              & 0.8047        & 0.8672        & 1.0000             & 0.9834        & 0.9916        \\
		                            & 6                          & 0.5826         & 0.5679         & 0.5752         & 0.9418         & 0.8247        & 0.8794        & 1.0000             & 0.9806        & 0.9902        \\
		\hline                            
	\end{tabular}
	\end{table*}
\end{center}

\begin{center}
	\begin{table*}[ht]
		\centering
		\caption{Data Comparison of Two Different Nodes of the IEEE 118-Bus System \\with Different Number of Communication Rounds (Under Strong Attacks)}
		\label{table4}
		\renewcommand{\arraystretch}{1.4}
		\begin{tabular}{ccccccccccc}
		\hline
		\multirow{2}{*}{Bus}         & \multirow{2}{*}{$R$}         & \multicolumn{3}{c}{CNN}                         & \multicolumn{3}{c}{LSTM}                      & \multicolumn{3}{c}{The Proposed Method}                      \\\cline{3-11}
		                           &                            & Precision           & Recall           & F1-score            & Precision           & Recall           & F1-score            & Precision           & Recall           & F1-score            \\
		\hline
		\multirow{4}{*}{55}         & 0                          & 0.7602         & 0.8475         & 0.8015         & 0.8776         & 0.8061         & 0.8403        & 1.0000          & 0.9042        & 0.9497        \\
		                            & 2                          & 0.9065         & 0.8883         & 0.8973         & 0.9962         & 0.8916           & 0.9410         & 1.0000          & 1.0000             & 1.0000             \\
		                            & 4                          & 0.9379         & 0.9086         & 0.9230          & 0.9991         & 0.9554         & 0.9767        & 1.0000          & 1.0000             & 1.0000             \\
		                            & 6                          & 0.9426         & 0.9120          & 0.9271         & 0.9978         & 0.9779         & 0.9877        & 1.0000          & 1.0000             & 1.0000             \\
		\hline
		\multirow{4}{*}{87}         & 0                          & 0.7981         & 0.7813         & 0.7896         & 0.9590          & 0.8582         & 0.9058        & 1.0000          & 1.0000             & 1.0000             \\
		                            & 2                          & 0.7983         & 0.7813         & 0.7897         & 0.9684         & 0.9232         & 0.9453        & 1.0000          & 1.0000             & 1.0000             \\
		                            & 4                          & 0.8656         & 0.7973         & 0.8300           & 0.9675         & 0.9393         & 0.9530         & 1.0000          & 1.0000             & 1.0000             \\
		                            & 6                          & 0.8671         & 0.7962         & 0.8302         & 0.9730          & 0.9434         & 0.9584        & 1.0000          & 1.0000             & 1.0000          \\
		\hline
	\end{tabular}
	\end{table*}
\end{center}

\vspace{-1.5cm}
\subsection{Detection Results}\label{Detection Results}
1) Performance evaluation metrics: In our experiments, we calculate the True Positive, False Positive, True Negative, and False Negative, where True Positive stands for the compromised measurement value is detected as the compromised data. Then we choose accuracy, precision, recall, and F1-score as the evaluation metrics for our output results. The specific formulas for each indicator can be found in the relevant literatures~\cite{12} and~\cite{20}.

\begin{figure}[ht]
	\centering
	\subfloat[]{\includegraphics[width=2.5in]{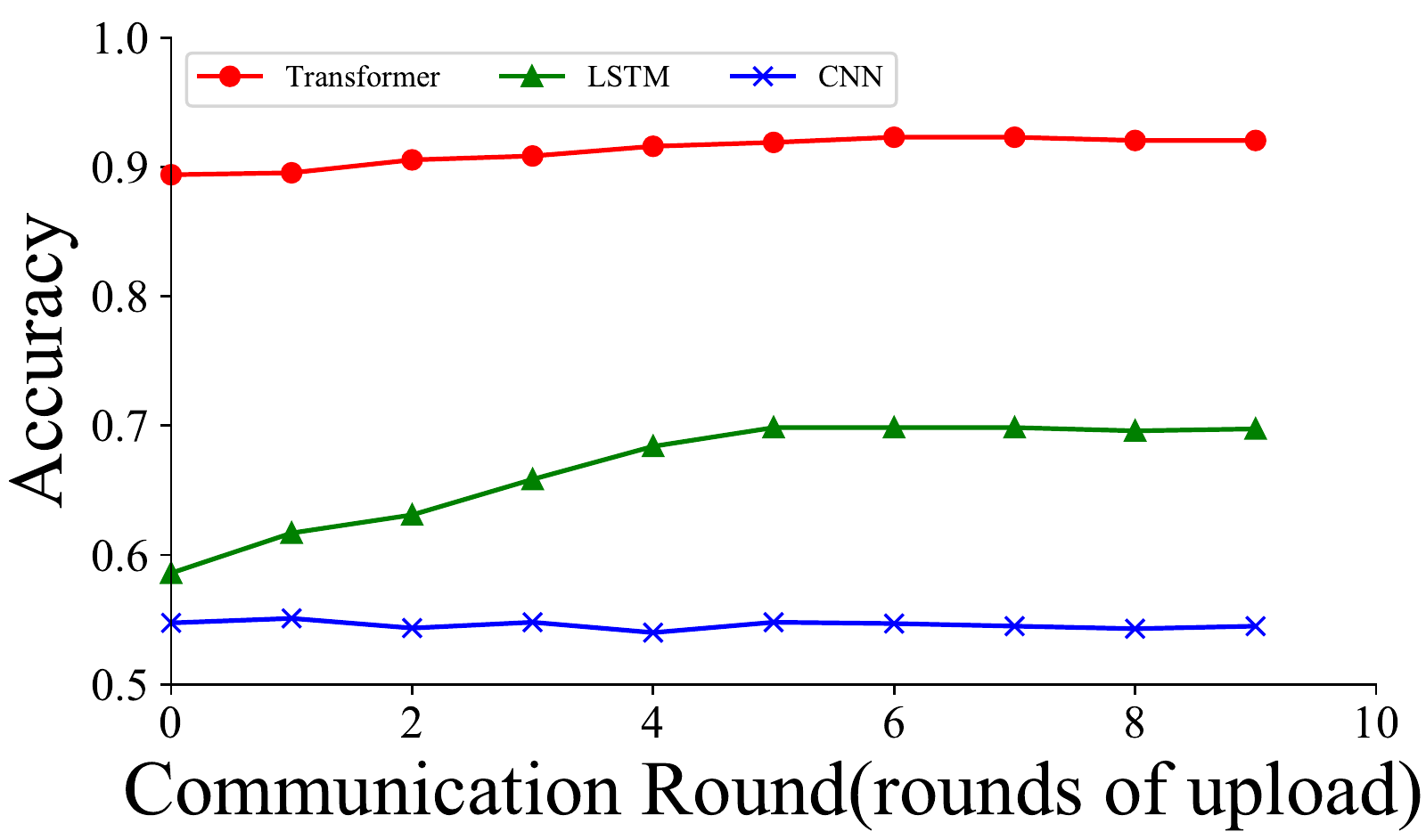}
		\label{fig_first_case1}
	}
	\hfil
	\subfloat[]{\includegraphics[width=2.5in]{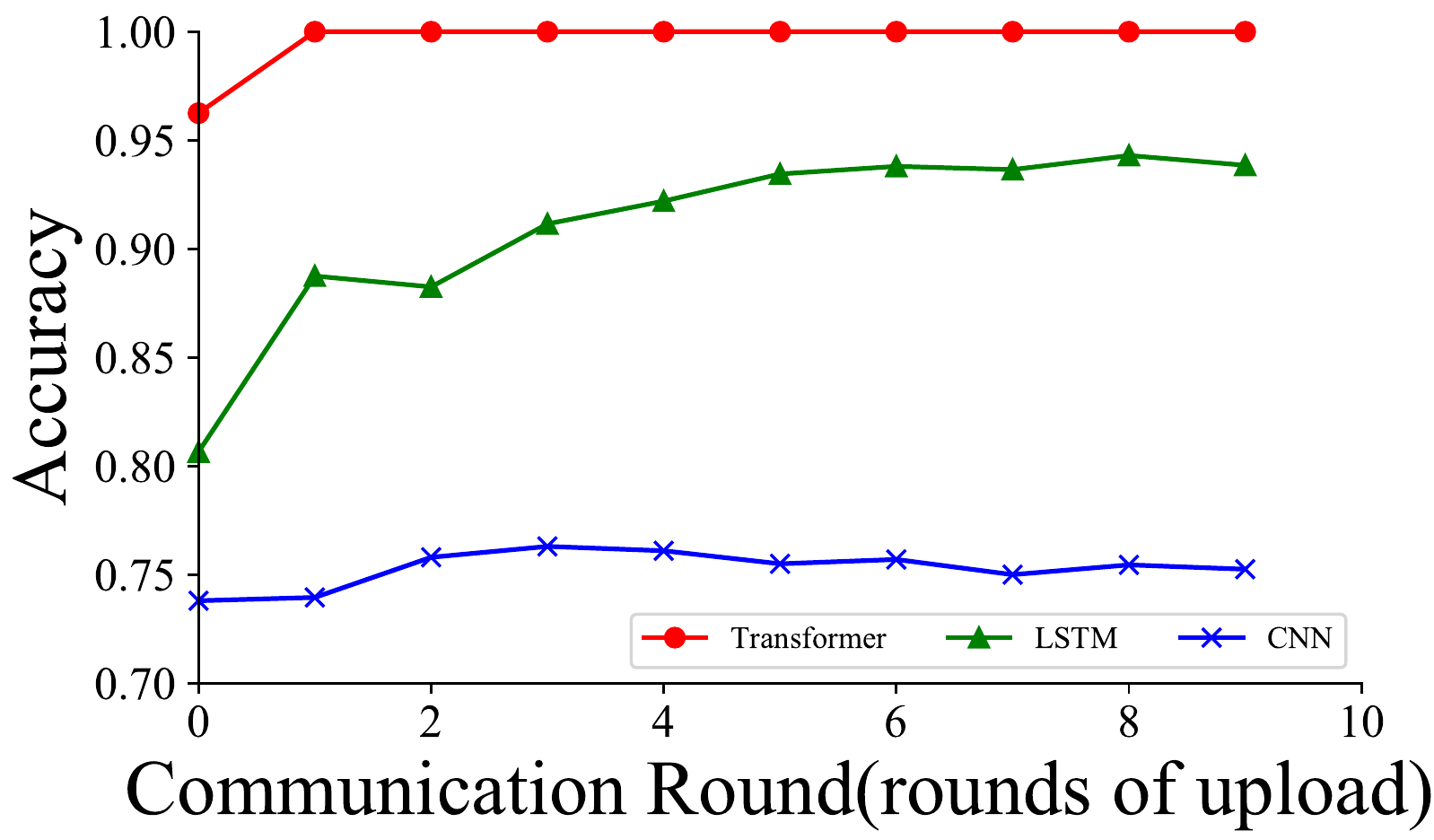}
		\label{fig_second_case1}
	}
	\caption{Comparison of  Accuracy of considered FDIA detection models with varying communication rounds in IEEE 14-bus system, bus 3. (a) Weak attacks. (b) Strong attacks.}
	\label{fig_sim1}
\end{figure}

2) The IEEE 14-bus system:~\reftab{table1} and~\reftab{table2} show the various evaluation metrics of the CNN, LSTM, and the proposed method in this paper for the IEEE 14-bus system under weak and strong attacks, respectively. We take bus 2 and bus 3 as examples to show the change of their metrics in different communication rounds ($R$).~\reffig{fig_first_case1} and~\reffig{fig_second_case1} show the change of accuracy with the number of communication rounds for the IEEE 14-bus system under weak and strong attacks, respectively, for the CNN, LSTM, and the proposed method in this paper, taking bus 3 as an example.

Under weak attacks, when $R$=6, accuracy, precision, recall, and F1-score are 0.9965, 0.9995, 0.9938, 0.9966 for bus 2, and accuracy, precision, recall, and F1-score are 0.9230, 0.9855, 0.8883, 0.9344 for bus 3, respectively. From the experimental results, it is clear that the proposed method is notably superior to the CNN and LSTM. As the communication round increases from 0 to 9 ($R$=0 means the model is trained locally without federated learning), the performance of each detection model is gradually enhanced and eventually stabilized; and the detection accuracy of the proposed method can reach more than 90\% for both strong and weak attacks. Furthermore, the accuracy of the proposed method always exceeds that of the CNN and LSTM. In addition, the proposed approach performs better under strong attacks than under weak attacks because strong attacks make the difference between normal data and compromised data more pronounced compared to weak attacks.


\begin{figure}[h]
\centering
\subfloat[]{\includegraphics[width=1.7in]{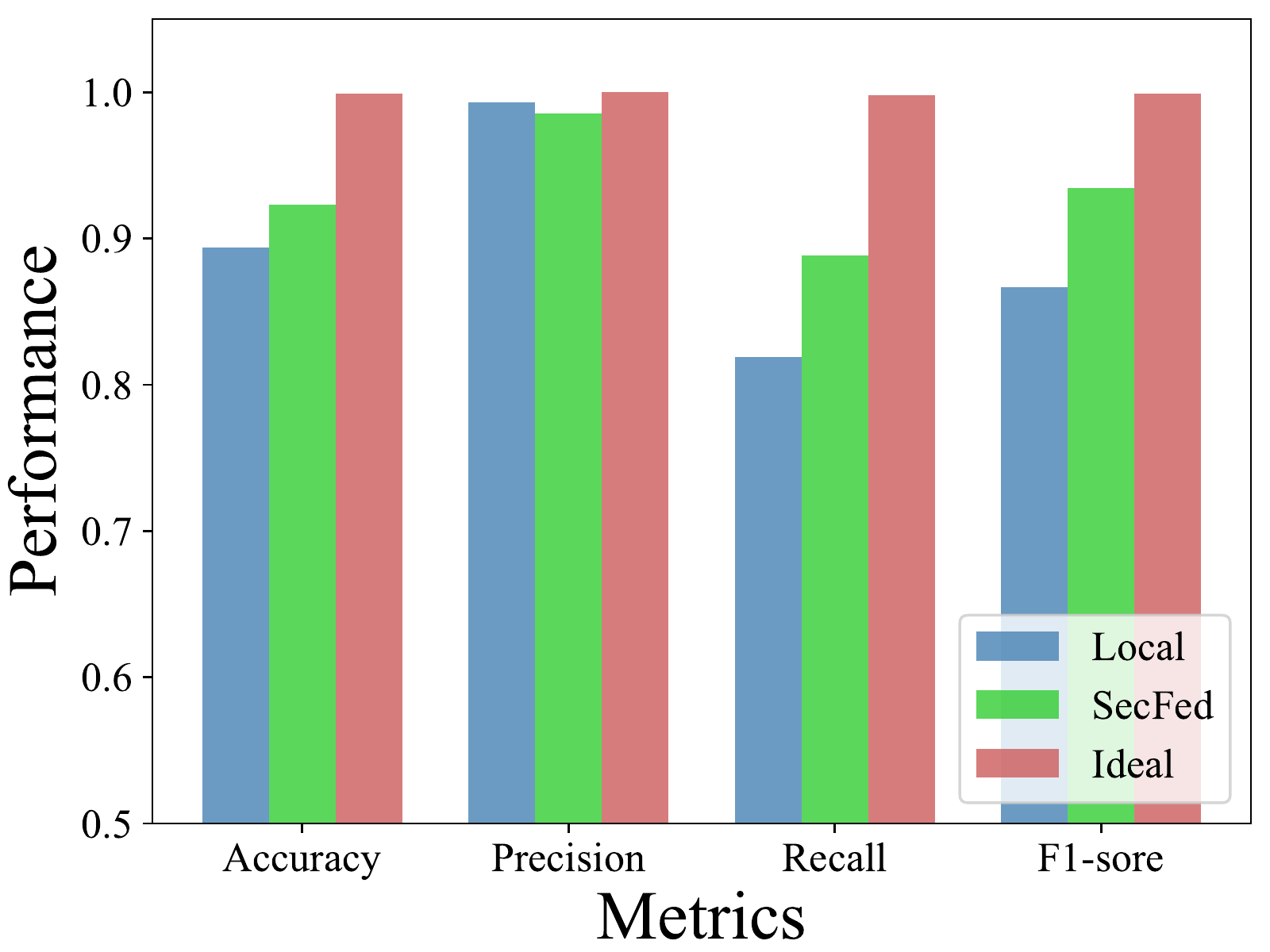}%
\label{fig_first_case2}}
\hfil
\subfloat[]{\includegraphics[width=1.7in]{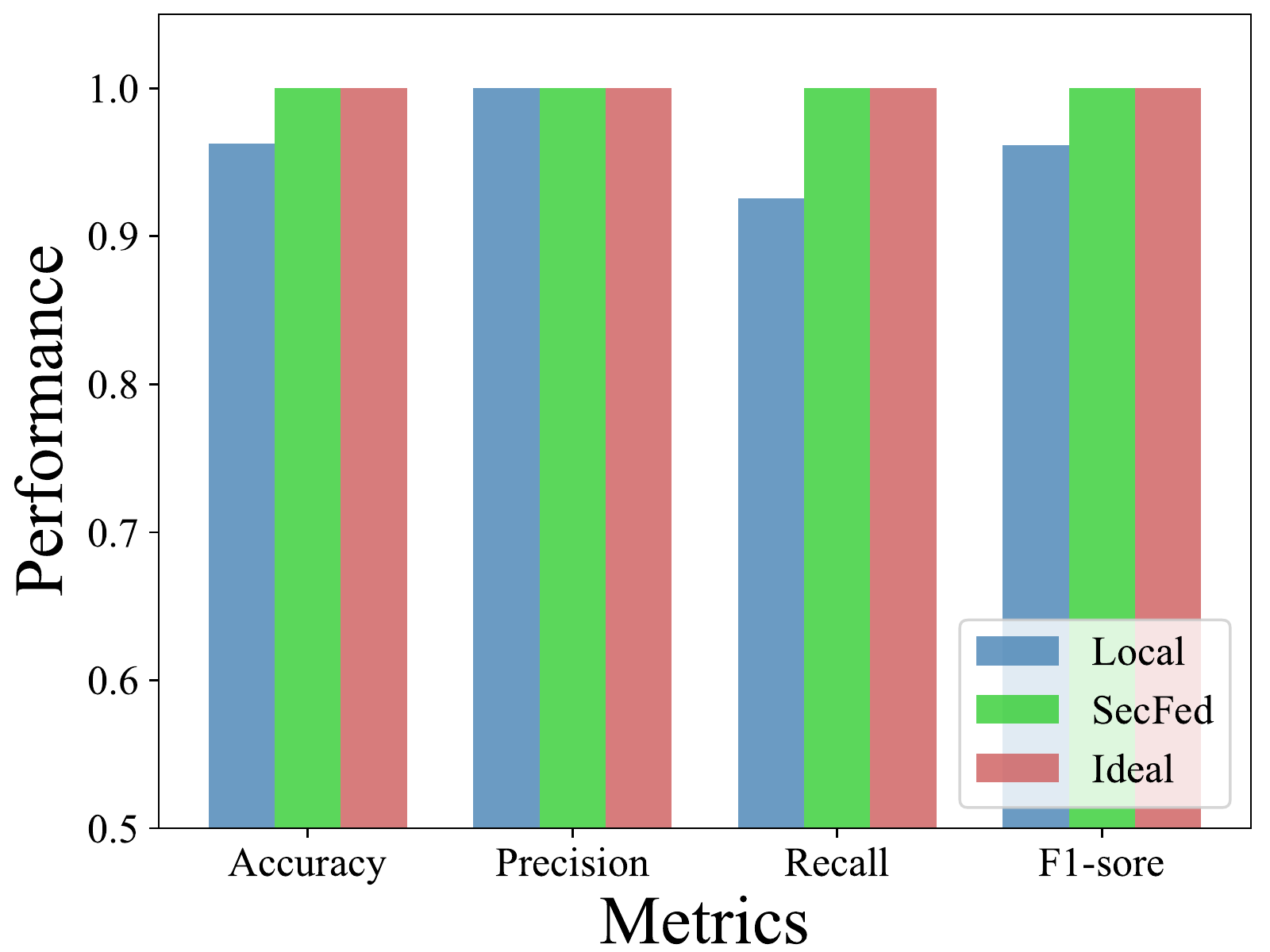}%
\label{fig_second_case2}}
\caption{Performance comparison of the local, ideal, and SecFed at bus 3 in the IEEE 14-bus system. (a) Weak attacks. (b) Strong attacks.}
\label{fig_sim2}
\end{figure}
To test the effectiveness of federated learning, we also conduct experiments about local model and ideal model, where local model means that each node uses only local data to train its own model without any communication with other nodes, and the ideal model means that the data of all nodes are pooled together and trained with a centralized model. \reffig{fig_sim2} shows the data differences between the above mentioned local model, federated model, and ideal model under weak and strong attacks. As seen in the figures, the accuracy, precision, recall, and F1-score of the local model are 0.9625, 1.0000, 0.9257, and 0.9614, respectively, while the accuracy, precision, recall, and F1-score of our proposed model are 1.0000, 1.0000, 0.9998, and 0.9999. It is obvious that the local model does not perform satisfactorily, while our method has a satisfactory performance. The performance of ours is approximately equivalent to that of the ideal model, but the SecFed scheme solves the problems existing in the ideal model,  such as communication delay, and privacy leakage. Therefore, compared with other deep learning methods, our model has the best performance in solving the FDIA detection problem in this study.

\begin{figure}[htbp]
	\centering
	\subfloat[]{\includegraphics[width=2.5in]{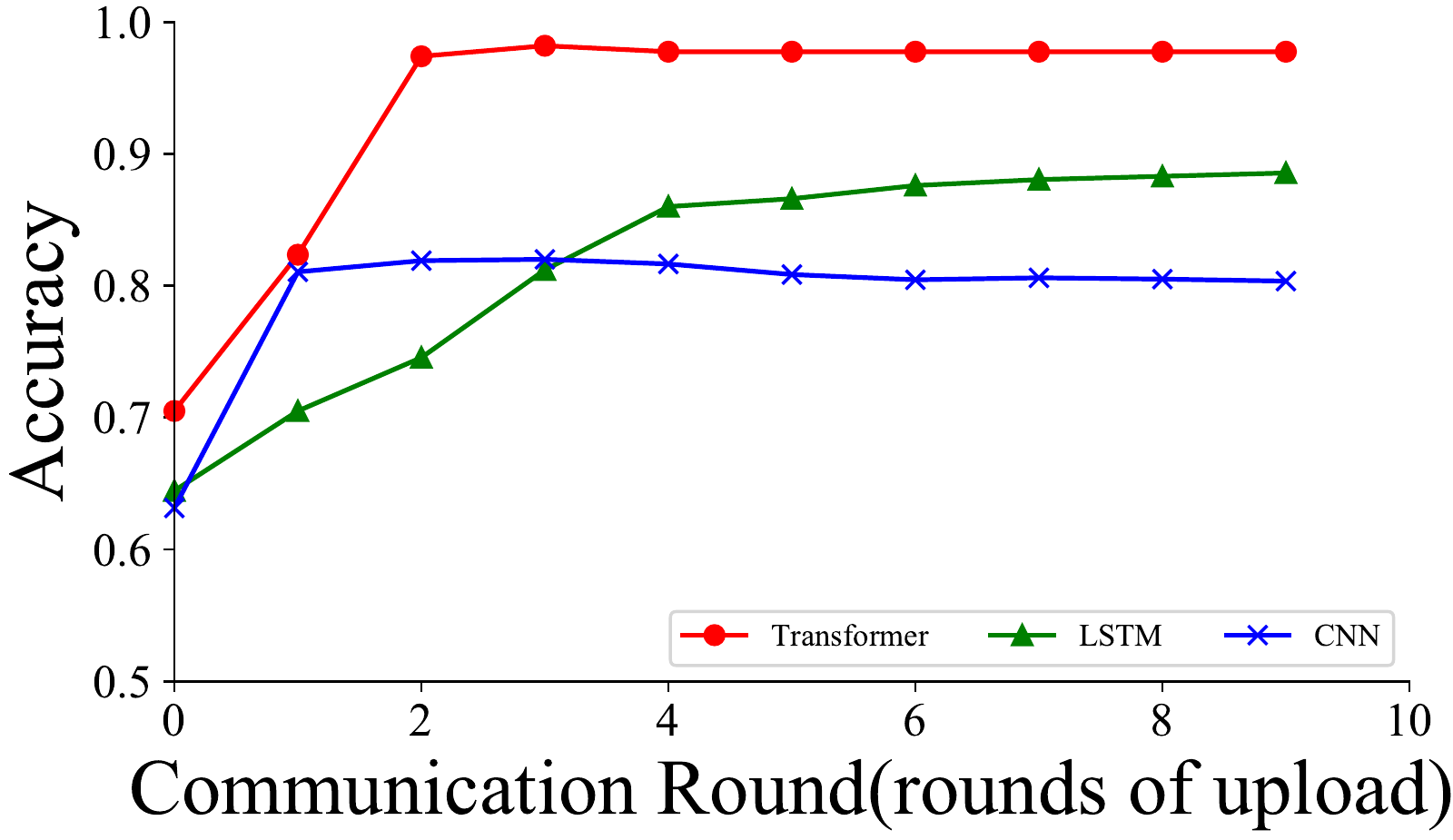}
		\label{fig_first_case3}
	}
	\hfil
	\subfloat[]{\includegraphics[width=2.5in]{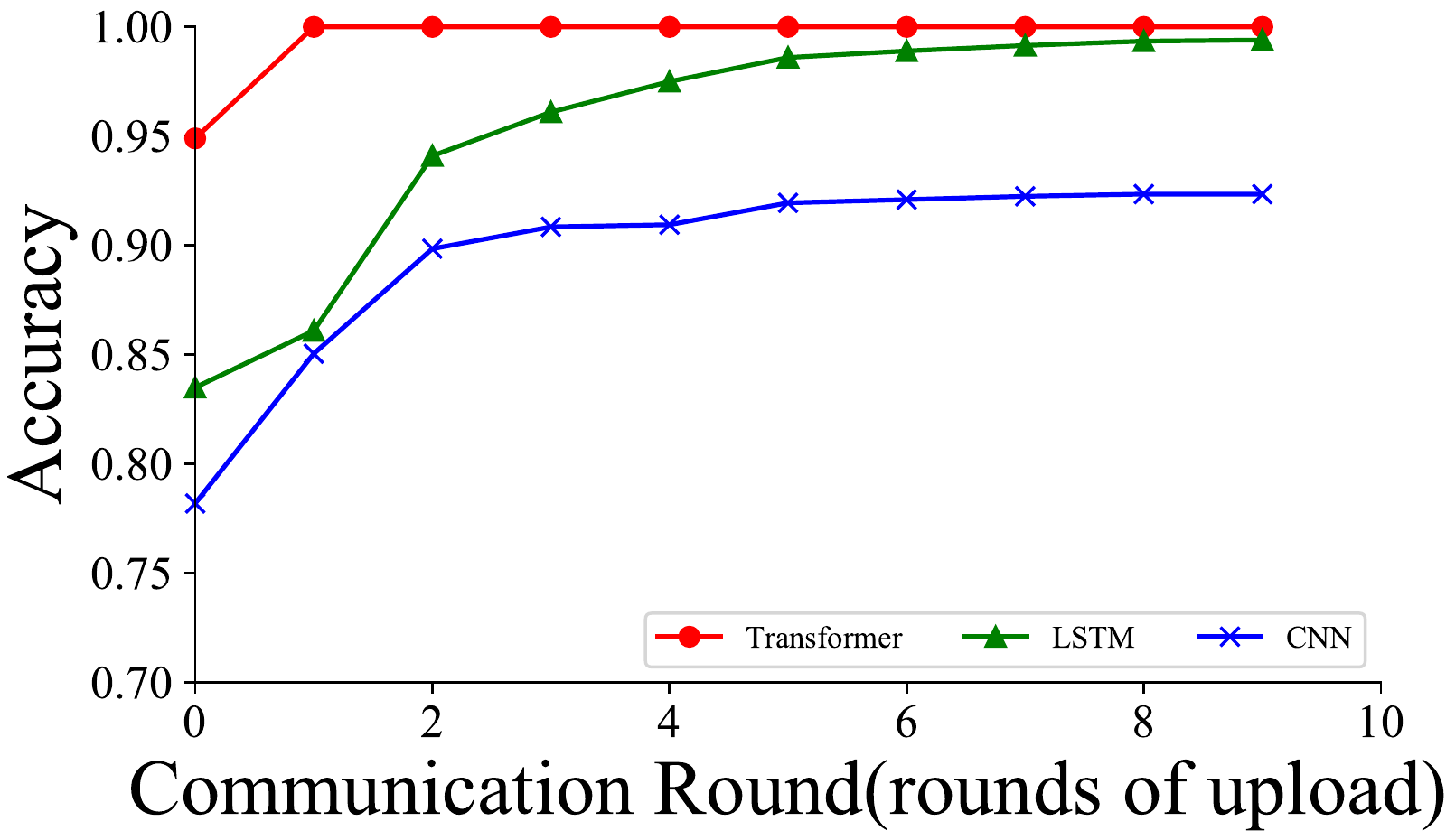}
	\label{fig_second_case3}
	}
	\caption{Comparison of the accuracy of considered FDIA detection models with varying communication rounds in IEEE 118-bus system, bus 55. (a) Weak attacks. (b) Strong attacks.}
	\label{fig_sim3}
\end{figure}

\begin{figure}[h]
\centering
\subfloat[]{\includegraphics[width=1.7in]{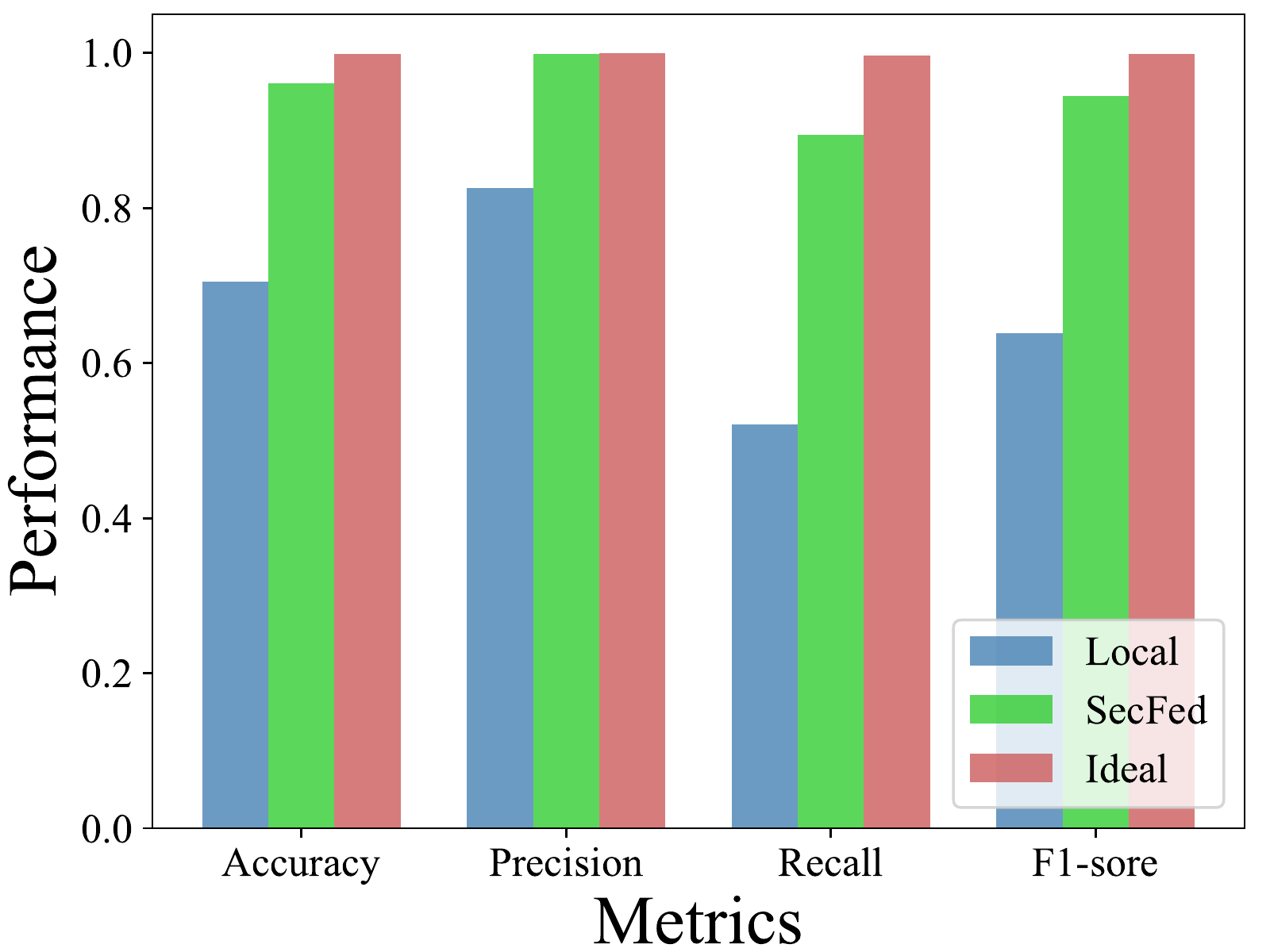}%
\label{fig_first_case4}}
\hfil
\subfloat[]{\includegraphics[width=1.7in]{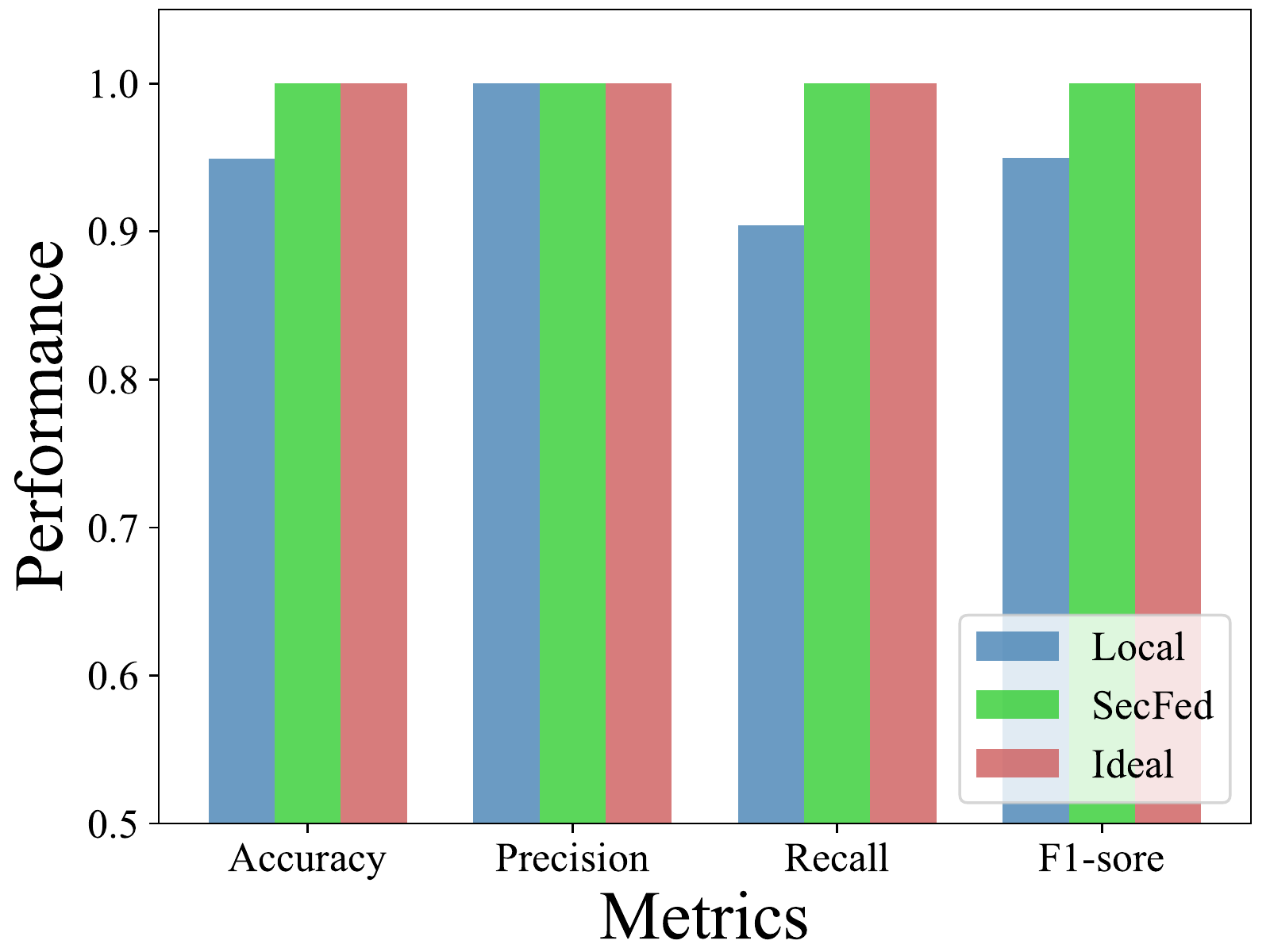}%
\label{fig_second_case4}}
\caption{Performance comparison of the local, ideal, and SecFed at bus 55 in the IEEE 118-bus system. (a) Weak attacks. (b) Strong attacks.}
\label{fig_sim4}
\end{figure}

3) The IEEE 118-bus system:~\reftab{table3} and~\reftab{table4} show the evaluation metrics of the CNN, LSTM and the proposed method in this paper for the IEEE 118-bus system under weak attacks and strong attacks, respectively. We take bus 55 and bus 87 as examples to show the change of the metrics of different models in different communication rounds.~\reffig{fig_sim3} shows the change of accuracy with the number of communication rounds for the IEEE 118-bus system under weak and strong attacks, for the CNN and LSTM, and the proposed method with bus 55 as an example. Similarly, we also conduct experiments concerning the local model and the ideal model.~\reffig{fig_sim4} shows the comparison of the results of each experiment.

From the results, it can be seen that the results on the IEEE 118-bus system are roughly similar to those on the IEEE 14-bus system, and the detection accuracy of the proposed model is significantly better than that of the CNN and LSTM.

\subsection{The Effect of Measurement Noises}\label{The Effect of Measurement Noises}
Due to the presence of noises in real power system measurement equipments~\cite{noise}, we design various levels of measurement noises and observe the impact of measurement noises. We add Gaussian noises to individual client data to represent the noises in the power system. In this experiment, we set the measurement noises to range from 1\% to 4\% of the true value.

\begin{figure}[h]
	\centering
	\includegraphics[width=3.5 in]{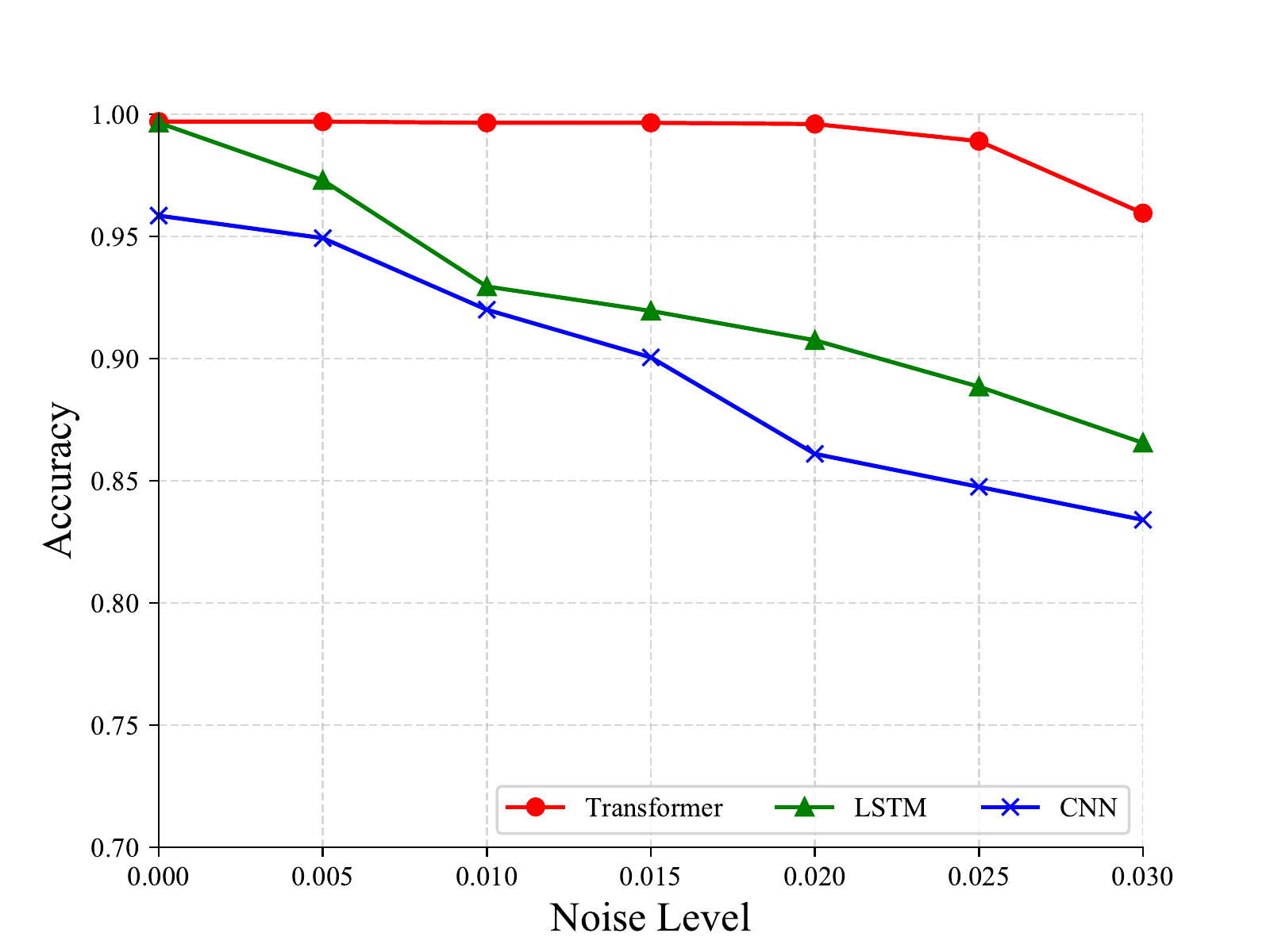}
	\caption{Performance of CNN, LSTM, Transformer under different noise levels.}
	\label{fig_8}
\end{figure}

\indent We take the accuracy of bus 4 as an example and compare the proposed method with the CNN and LSTM in this paper. The specific results are displayed in~\reffig{fig_8}. From the figure, it is easy to see that the accuracys of all models decrease as the noise level increases. This is because with the increasing noise levels, normal data and compromised data become more difficult to distinguish as the unnoised data is more likely drowned in the noises. Moreover, it is obvious from the figure that our proposed method performs significantly better than the CNN and LSTM under the influence of noises, therefore it can be concluded that our model has better robustness than other alternatives..

\section{Conclusion}
Aiming at false data injection attacks faced by smart grids, a FDIA detection method combining secure federated learning with Transformer is proposed in this paper. To the authors' knowledge, this is the first work to leverage secure federated learning for detecting FDIA in smart grid. First, we build a Transformer-based FDIA detection model for local training of each node. Then, we introduce a federated learning framework to enable all nodes collaboratively train a detection model while preserving the privacy of all the local training data. In addition, we combine the Paillier cryptosystem with the federated learning framework to construct secure federated learning, which effectively protects the privacy of federated learning during training. Extensive experimental results indicate that our Transformer-based detection model significantly outperforms conventional deep learning algorithms and the proposed secure federated learning approach has unique advantages in protecting data privacy and reducing communication overhead than centralized detection methods.

In the future, we will extend our method to handle the detection of multi-cyber attacks, including but not limited to denial-of-service attacks and replay attacks. This work assumes that an attacker has all system information during attacks, while more realistic scenarios should consider cyber attacks with incomplete information. It's also interesting to determine the hyper-parameters of the presented approach by using automated machine learning~\cite{wangruinong}.

{\appendix
Paillier cryptosystem is an additive homomorphic key cryptosystem for securing federated learning data, and its key mechanism consists of the following four parts:

1. ${KeyGeneration}$: Trustee randomly chooses two large prime numbers \emph{p} and \emph{q} and calculates their product $n$ as well as $\lambda$ ($\lambda$ is the least common multiple of \emph{p} - 1 and \emph{q} - 1). Then, pick a random integer $g \in Z_{{n^2}}^*$,and $g$ must satisfy
\begin{equation}
	\begin{aligned}
		\gcd (L({g^\lambda }\bmod {n^2}),n) = 1.
		\label{XX16}
	\end{aligned}
\end{equation}
}

The final calculation is made $\mu={(L({g^\lambda }\bmod {n^2}))^{ - 1}}$, where the function $L(x) = (x - 1)/n$, and the function $\gcd ( \cdot )$ is used to calculate the maximum common divisor of two numbers. $Z_{{n^2}}^{}$ is the set of integers less than ${n^2}$, and $Z_{{n^2}}^*$ is the set of integers that are mutually exclusive with ${n^2}$ in $Z_{{n^2}}^{}$ .
We choose $(n, g)$ as the public key(${PK}$), and $(\lambda ,\mu )$ as the private key (${CK}$) of each client and trustee distributes them to each client.

2. ${Encryption}$($w, {PK}$): To ensure that the encrypted number is a positive integer, we preprocess the weights. Define a function $\hat x = f(x) = {10^8} \times (x + s)$, where $s$ is greater than the absolute value of the minimum value of the weights. The weights of the local client can be transformed to positive integer ${\hat w_{td}} \in {Z_n}$ by calculation ${\hat w_{td}} = f({w_{td}})$ . After randomly selecting a positive integer $r \in Z_n^*$, we encrypt the model weights as follows:
\begin{equation}
	\begin{aligned}
		{c_{td}} = {g^{f({w_{td}})}} \cdot {r^n}\bmod {n^2},
		\label{XX17}
	\end{aligned}
\end{equation}
where $\emph{t} \in \{1, 2, \cdots, T\}$, $\emph{d} \in \{1, 2, \cdots, N\}$. \emph{T} represents the number of clients, and \emph{N} denotes the number of weights in each client.

3. ${Aggregation}$($c_{td}$): Each client uploads the encrypted weight ${c_{td}}$ to server, then the cloud server aggregates the received parameters by
\begin{equation}
	\begin{aligned}
		{a_d} = {c_{1d}} \cdot {c_{2d}} \cdot \cdots \cdot {c_{Kd}} = \prod\limits_{t = 1}^K {{c_{td}}}.
	\end{aligned}
\end{equation}

4. ${Decryption}$($(a_{d}, {CK}$): The server distributes the aggregated weights to each client, and each client decrypts the received ciphertext by
\begin{equation}
	\begin{aligned}
		{m_d} &= L({a_d}\bmod {n^2}) \cdot \mu \bmod n
		\\&= L(a_d^\lambda \bmod {n^2}) \cdot L{({g^\lambda }\bmod {n^2})^{ - 1}}\bmod n.
	\end{aligned}
\end{equation}

Afterwards, the average value of the decrypted parameters with respect to the client will be calculated ${\bar m_d} = {m_d}/K$. The weights which are updated are calculated by 
\begin{equation}
	\begin{aligned}
		w{'_{td}} = {f^{ - 1}}({\bar m_d}) = {10^{ - 8}} \cdot {\bar m_d} - s.
	\end{aligned}
\end{equation}

\bibliographystyle{ieeetr}
\bibliography{ref.bib}

\end{document}